# Unified theory of the direct or indirect bandgap nature of conventional semiconductors

Lin-Ding Yuan[1,2], Hui-Xiong Deng[1,2], Shu-Shen Li[1,2,3], Jun-Wei Luo[1,2,3*], and Su-Huai Wei[4*]

[1]*State key laboratory of superlattices and microstructures, Institute of Semiconductors, Chinese Academy of Sciences, Beijing 100083, China*

[2]*College of Materials Science and Opto-Electronic Technology, University of Chinese Academy of Sciences, Beijing 100049, China*

[3]*Beijing Academy of Quantum Information Sciences, Beijing 100193, China*

[4]*Beijing Computational Science Research Center, Beijing 100193, China*

*Email: jwluo@semi.ac.cn; suhuaiwei@csrc.ac.cn

**Abstract:**

Although the direct or indirect nature of the bandgap transition is an essential parameter of semiconductors for optoelectronic applications, the understanding why some of the conventional semiconductors have direct or indirect bandgaps remains ambiguous. In this Letter, we revealed that the existence of the occupied cation *d* bands is a prime element in determining the directness of the bandgap of semiconductors through the *s-d* and *p-d* couplings, which push the conduction band energy levels at the X- and L-valley up, but leaves the Γ-valley conduction state unchanged. This unified theory unambiguously explains why Diamond, Si, Ge, and Al-containing group III-V semiconductors, which do not have active occupied *d* bands, have indirect bandgaps and remaining common semiconductors, except GaP, have direct bandgaps. Besides *s-d* and *p-d* couplings, bond length and electronegativity of anions are two remaining factors regulating the energy ordering of the Γ-, X-, and L-valley of the conduction band, and are responsible for the anomalous bandgap behaviors in GaN, GaP, and GaAs that have direct, indirect, and direct bandgaps, respectively, despite the fact that N, P, and As are in ascending order of the atomic number. This understanding will shed light on the design of new direct bandgap light-emitting materials.

Whether a semiconductor has a direct or indirect bandgap is of fundamental importance to its optoelectronic applications [2, 3]. If the conduction band minimum (CBM) occurs at the same point in k-space as the valence band maximum (VBM), which is usually at the center (Γ-point) of the Brillouin zone for conventional semiconductors, then the energy gap is referred as direct bandgap, otherwise as indirect bandgap [3]. If a semiconductor has a direct bandgap and the electric dipole transition from VBM to CBM is allowed, the electron-hole pairs will recombine radiatively with a high probability. As a result, high-quality direct bandgap semiconductors, such as GaAs and InP, are used to make high efficient light emitters. They are essential materials for lasers, light emitting diodes (LEDs) and other photonic devices [4, 5]. Whereas, in indirect bandgap semiconductors, such as Si and Ge, optical transitions across an indirect bandgap are not allowed, and, thus, these materials are not efficient light emitters. Despite the paramount importance of the direct or indirect nature of the bandgap transition for conventional semiconductors, which have been extensively studied in past seven decades, the understanding of the formation of their direct or indirect bandgaps remains ambiguous.

In spite of the extensive utilizing of semi-empirical and first-principles methods to correctly reproduce the experimentally measured band structures for semiconductors, the simple nearest-neighbor tight-binding (TB) theory is more straightforward to gain insight into the formation of the band structures because of its intuitive simplicity[6]. However, this simple TB model fails to reproduce some important band structure features, such as bandgap nature for indirect semiconductors[3, 7]. Although the introduction of additional unphysical parameters can cure the flaw of the simple $sp^3$ TB model[8-11], it loses the advantage of its intuitive simplicity and thus is unlikely to uncover the origin of the direct and indirect bandgap natures of semiconductors. The poor understanding impedes the design of new direct bandgap light-emitting materials. Specifically, Si is ubiquitous in the electronics industry but is unsuitable for optoelectronic applications because it has an indirect bandgap. In the past five decades, numerous ideas have been offered but failed to transform Si into an efficient light emitter [12, 13] utilizing various modalities of symmetry reduction, including the use of porous silicon [14, 15], invoking alloy-induced luminescence [16, 17], the method of low-symmetry allotropes of silicon [18-20], and diamond-like (III−V)−(IV) alloys[21, 22].

Lack of fundamental understanding of the mechanism controlling the indirect bandgap nature of Si might be the main reason for the difficulty of developing Si-based direct bandgap materials.

Here, we reveal that the occupied cation *d* bands, which were neglected in previous models of the TB approach [6-11], play a prime role in forming the direct bandgap of semiconductors via the *s-d* and *p-d* couplings. These couplings repel the conduction band energy levels of X- and L-valley up but leave the Γ-valley intact. From group-IV through group III-V to group II-VI semiconductors, the occupied cation *d* orbitals become closer in energy to the anion *s* and *p* orbitals, leading the *s-d* and *p-d* coupling to be strongest in group II-VI semiconductors, and hence all their bandgaps being direct. The either lacking or low-lying of the occupied *d* orbitals in cations of Diamond, Si, Ge, and Al-containing group III-V semiconductors is responsible for their nature of indirect bandgap.

We at first examine the nature of bandgaps of all conventional group-IV elemental, and group III-V and group II-VI compound semiconductors, which are the semiconductors of practical interest for information technology[3, 5, 23]. Fig. 1 shows that all group II-VI compound semiconductors and the majority of group III-V compound semiconductors, except Al-containing compounds and GaP, have a direct bandgap, whereas all group-IV elemental materials, except grey-Sn, are indirect bandgap [23, 24]. We note that the cations of all group-II elements (Zn, Cd, and Hg), group-III elements Ga and In, and group-IV elements Ge and Sn contain occupied *d* orbitals, which, however, are absent in the remaining group-III element Al and group-IV elements C and Si. Because cation elements of all direct bandgap semiconductors encompassing occupied *d* orbitals, whereas all semiconductors made of cations without occupied *d* orbitals have indirect bandgaps, it strongly suggests that the occupied cationic *d*-orbitals play a central role in determining the directness of the bandgap for conventional semiconductors. However, the filled *d* shells, if they exist, are often treated as core or semi-core shells and are usually neglected in the description of the band structures for conventional semiconductors in early studies [6-11]. To uncover the role of the cationic *d* orbitals in the formation of bands, we carried out the theoretical analysis of the band structures of diamond elements and zinc-blende (ZB) compounds relying on the perturbation theory along with the symmetry analysis. For simplicity, here we study only the band structures of compounds in the zinc-blende structure, GaN and ZnO, they prefer to be in the hexagonal wurtzite (WZ) structure under ambient conditions. The relationship between the bandgaps of ZB and WZ structures are well studied [25]. For example, if the band gap is direct in ZB structure such as for GaN and ZnO, it is also direct in WZ structure; if the band gap is indirect in ZB structure such as for AlN, it could still be direct in WZ structure due to the average of the X and L point wavefunctions.

The relative energy positions between the Γ-valley, X-valley, and L-valley in the lowest conduction band determine the direct or indirect nature of bandgap in ZB semiconductors since the VBM occurs exclusively at the Γ point in all group II-VI, group III-V, and group IV semiconductors. As given in Table I [1], in the zinc-blende structure, *the CB edge at the Γ point* (Γ-valley) transforms according to the $\Gamma_1$ irreducible representation, the *VB edge* transforms according to the $\Gamma_{15}$ irreducible representation, and the atomic *d* orbitals belong to the $\Gamma_{15}$ and $\Gamma_{12}$, respectively. Since the *d* orbitals have the same $\Gamma_{15}$ irreducible representation as the p-like VB edge state, the coupling between *p* and *d* orbitals at the Γ point could be quite significant. However, the *s-d* coupling is forbidden because the atomic *d* orbitals have no common irreducible representations with the *s*-like CB Γ-valley state. Therefore, the existence of the occupied *d* orbitals will have a significant influence on the formation of bandgap by pushing the VBM up and leaving the CB Γ-valley intact. The *CB edge at the X point* (X-valley) transforms according to the $X_1$ (or $X_3$) irreducible representation of the $D_{2d}$ wavevector group, and mainly derived from atomic *s* and *p* orbitals, whereas, five *d* orbitals belong to the $X_1$, $X_2$, $X_3$, and $X_5$ irreducible representations, respectively. Therefore, at the X point, the *d* orbital state can couple to the CB X-valley. The mostly *s-like CB edge at the L point* (L-valley) has the $L_1$ representation of the $C_{3v}$ wavevector group, whereas, the five *d* orbitals belong to the $L_1$ and two $L_3$ representations, respectively. Same as the X point, the *d* orbital state can couple to the CB L-valley. Subsequently, the existence of the occupied d orbitals will repel the CB X-valley and L-valley up due to the symmetry allowed *s-d* coupling and *p-d* coupling at the X and L points, but will not affect the CB Γ-valley owing to its lack of atomic *p* orbitals plus symmetry forbidden *s-d* coupling at the Γ point (see supplementary for diamond structure). Such *s-d* coupling at the X and L points is evidenced by the fact that, due to the low-lying *s* orbital energy of nitrogen, the *s-d* coupling away from the Γ point is so strong in GaN and InN that leads to two s-like peaks observed in their photoemission near the bottom of the valence band[26]. These *s-d* coupling not only makes significant influences on the band structure near the bottom of the valence band but should also largely affect the conduction band at both X and L points. Unfortunately, the latter has yet uncovered even though a remarkable amount of the *d* character was found thirty years ago in the CB X- and L-valley in conventional semiconductors [27].

To illustrate the effect of the occupied *d* orbitals on the nature of bandgap, we examined the energy level shifting of the CB Γ-, X-, and L-valley caused by the *s-d* and *p-d* coupling due to the existence of the occupied *d* shells, which were neglected previously, as schematically shown in Fig. 2(a). We took GaAs as the prototype to demonstrate the suggested unique role of the occupied cation *d* orbitals (the anion *d* orbitals are so low in energy that is negligible compared with the cation *d* orbitals). Fig. 2(b) shows the first-principles calculated GaAs band structure using the density functional theory (DFT) based on the modified Becke-Johnson (mBJ) exchange potential in combination with the generalized gradient approximation (GGA) correlation (mBJ-GGA). One can see that the Ga 3*d* bands with narrow bandwidths occurring about 15 eV below the VBM. The interaction between *d* bands and the *p*-like VBM via *p-d* coupling leads to a significant amount of the *d* character in the VBM (inset to Fig. 2(b)). Since the *s-d* and *p-d* couplings are allowed at X and L points, the Ga 3*d* bands repel the X- and L-valley up in a significant amount of energy as evidenced by the incorporation of finite *d* component in both $X_{1c}$ and $L_{1c}$ states, in addition to expected dominated *s* and *p* components. Whereas, inset to Fig. 2(b) shows the vanishing of the *p* and *d* characters in the $Γ_{1c}$ state, confirming it to be purely an anti-bonding state of Ga 4*s* and As 4*s* and immune to the existence of the Ga 3*d* bands.

To examine the effect of *s-d* and *p-d* coupling on the band edges, we artificially pull the Ga 3*d* bands down to modify the *s-d* and *p-d* hybridizations. An adjustable Coulomb-U acting on Ga 3*d* orbitals is used as an effective knob to tune the energy position of Ga *d*-levels by introducing Hubbard-type interactions into the DFT (DFT+U). This method has been widely used to correct the underestimated DFT bandgaps by pulling the VBM down in energy through a *p-d* coupling and leaving the CBM at Γ intact[28, 29]. Here, we applied this method to investigate the impact of the Ga 3*d* on the energies of the CB Γ-, X-, and L-valley. Fig. 2(b) shows the change of the energy positions of X-, L-valley, and VBM as varying the energy position of the Ga 3*d* bands, regarding the Γ-valley as an ideal reference level since, to the lowest order, it is free from the change of the *d* bands. As we pull the Ga 3*d* bands down, the X- and L-valley also move down in energy but in different rates, and, finally, the GaAs bandgap becomes indirect, demonstrating the importance of the energy position of the occupied cation *d* bands in determining the nature of bandgap unambiguously.

The energy position of the cation *d* bands relative to the (anion *p* dominated) VBM in conventional semiconductors depends mainly on the energy separation between the outermost anion *p* shell and the cation *d* shell, which decreases in the sequence of group-IV, III-V, II-VI semiconductors as shown in Fig. S1. Table SI gives energy positions of the cation *d* bands relative to the VBM predicted by the mBJ-GGA calculations in comparison with experimental data [30-33] for those conventional semiconductors possessing occupied cation *d* orbitals. They are about -10 eV for group II-VI compounds, -18 eV for group III-V compounds, and -25 eV for group-IV elemental semiconductors, respectively. Going from group-IV through group III-V to group II-VI semiconductors, the shallower cation *d* bands lead to stronger *s-d* and *p-d* couplings and hence larger repulsion of the X- and L-valley by the low-lying occupied cation *d* bands. In compared with GaAs, the enhanced *s-d* and *p-d* couplings in the group II-VI semiconductors repel the X- and L-valley more and lead to the bandgap more direct. Whereas, in group IV Ge, the low-lying Ge 4*d* bands result in weak *s-d* and *p-d* couplings so that its bandgap becomes indirect. In the group IV Diamond and Si and Al-based group III-V semiconductors, because the cation *d* orbitals are above the CB edges rather than in the occupied valence bands, the *s-d* and *p-d* couplings push CB X- and L-valley down instead of repelling them up, and subsequently make their bandgap indirect.

Given that Ga possesses occupied 3*d* orbitals, the GaN and GaAs bandgaps are as expected direct. However, GaP sitting in the middle between them is an indirect gap semiconductor. This abnormal bandgap behavior indicates that besides the primary *s-d* and *p-d* couplings, other factors are also playing roles in determining the order of Γ-, X- and L-valley in the lowest CB. We notice that the lattice constants of zinc-blende GaN, GaP, and GaAs are 4.531, 5.451, and 5.653 Å, respectively[24]. The bond length of GaP is ~3.6% smaller than that of GaAs. Fig. 4 shows that, in both GaP and GaAs, increasing the lattice constant (or expanding the volume) will raise the energy level of the X-valley and lower the Γ-valley energy substantially. The L-valley follows the Γ-valley but at a much smaller rate and, thus, often sits in between Γ- and X-valley in conventional semiconductors. This phenomenon is owing to that the X-valley has a positive deformation potential and Γ-valley has a more significant magnitude of deformation potential than that of the L-valley although both possess negative deformation potentials. If we stretch the lattice of GaP to equal to that of GaAs, the GaP bandgap would become direct. On the other hand, GaAs undergoes a direct-to-indirect bandgap transition as we compress the lattice of GaAs towards that of GaP. These demonstrate that besides *s-d* and *s-p* couplings, the bond length also play role in determining the nature of bandgap. Semiconductors having larger lattice prefer to become more direct in the bandgap.

Following the above discussion, we would expect the bandgap of GaN to be indirect since the bond length of GaN is much smaller than GaP and GaAs. However, GaN is a classical direct gap semiconductor. Fig. 4 shows that the deformation potentials of GaN are substantially different from that of GaP and GaAs; the energy level of the X-valley is insensitive to the varying of the lattice constant with an inconsiderable negative deformation potential and the Γ-valley drops at the same rate as that of the L-valley as increasing the lattice constant. AlN and InN share this exotic behavior as GaN [34]. This unusual behavior of the group-III nitrides is due to that Nitrogen is among the most electronegative elements and much more electronegative than P and As elements. Fig. S1 shows that the energy level of the N 2*s* is 5.94 and 5.71 eV lower than that of the P 3*s* and As 4*s*, respectively. The low-lying N 2*s* orbital is far away from the Ga 3*s* orbital leading to a weak *s-s* coupling in GaN according to the TB model [6] (see supplementary for details), although it has a much smaller bond length than that of GaP and GaAs. Weak *s-s* coupling results in the energy level of the Γ-valley, which is the anti-bonding state of cation *s* and anion *s* orbitals, being slightly above the Ga 3*s* level and is even lower than that of GaP and GaAs. Although the low-lying of the N 2*s* orbital also reduces the *s-p* coupling and thus lowers the energy level of the X-valley, the reduction in energy is much less than the Γ-valley due to the X-valley has the lower bound limited by the atomic Ga 3*p* level [6]. Consequently, GaN becomes a direct bandgap semiconductor (see Fig. 1(b)). Thus, more electronegativity in anions will also make the bandgap of semiconductors more direct, which is most significant in semiconductors containing O or N. Our analysis above indicates that despite size of the atoms and electronegativity of the anions can play some roles in determining the directness of the band gap, especially for the boundary Ga compounds which has relatively deep 3$d$ orbitals, the coupling of the occupied cation $d$ bands and unoccupied $s,p$ states plays the prime role in determining the bandgap nature as manifested by the indirect gap of ZB AlN.

In summary, we have presented a unified theory for understanding the direct or indirect nature of bandgap in group II-VI, group III-V, and group IV semiconductors unambiguously. We found that the occupied cation $d$ bands play a prime role in forming the direct or indirect bandgap of conventional semiconductors via the $s$-$d$ and $p$-$d$ coupling with the states of the CB X- and L-valley, which remarkably pushes their energy levels up, but leaves the $\Gamma$-valley unchanged. From group-IV through group III-V to group II-VI semiconductors, the occupied cation $d$ orbitals become closer in energy to the anion $s$ and $p$ orbitals, leading the $s$-$d$ and $p$-$d$ coupling to be most active in group II-VI semiconductors, and hence all their bandgaps being direct. The either lacking or low-lying of the occupied $d$ orbitals in cations of Diamond, Si, Ge, and Al-containing group III-V semiconductors explains their nature of indirect bandgap. This understanding will shed light on the design of new direct bandgap light-emitting materials.

**Acknowledgments**

J.W.L. was supported by the National Natural Science Foundation of China (NSFC) under Grant No. 61474116 and the National Young 1000 Talents Plan. S.H.W. was supported by the NSFC under Grant No. 51672023, 11634003, and U1530401 and the National Key Research and Development Program of China under Grant No. 2016YFB0700700.

TABLE I. The point group of the wave-vector at Γ, X, and L points in the zinc-blende structure and the corresponding irreducible representations of atomic *s*, *p*, and *d* orbitals as well as semiconductor conduction (CBE) and valence (VBE) band edges under these point groups [1].

| k-point | G(k) | *CBE(k)* | *VBE(k)* | *s* | *p* | *d* |
|---|---|---|---|---|---|---|
| Γ | $T_d$ | $\Gamma_1$ | $\Gamma_{15}$ | $\Gamma_1$ | $\Gamma_{15}$ | $\Gamma_{15} \oplus \Gamma_{12}$ |
| X | $D_{2d}$ | $X_1$ or $X_3$ | $X_5$ | $X_1$ | $X_3 \oplus X_5$ | $X_1 \oplus X_2 \oplus X_3 \oplus X_5$ |
| L | $C_{3v}$ | $L_1$ | $L_3$ | $L_1$ | $L_1 \oplus L_3$ | $L_1 \oplus L_3 \oplus L_3$ |

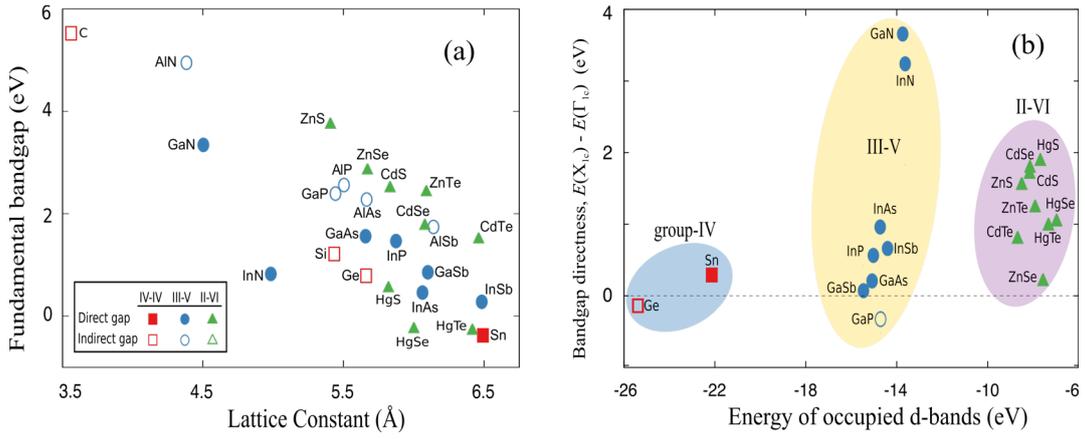

FIG 1. (a) Fundamental bandgaps of group II-VI, group III-V, and group IV semiconductors as a function of their lattice constants a. Filled symbols represent direct bandgap semiconductors and open symbols for indirect bandgap semiconductors. On the other hand, the squares indicate the group IV semiconductors, the circles for group III-V semiconductors, and the triangles for group II-VI semiconductors. (b) The energy difference between levels of the CB X-valley and Γ-valley as a function of the energy of the cation d bands (relative to the VBM) for semiconductors consisting of cations having occupied d orbitals.

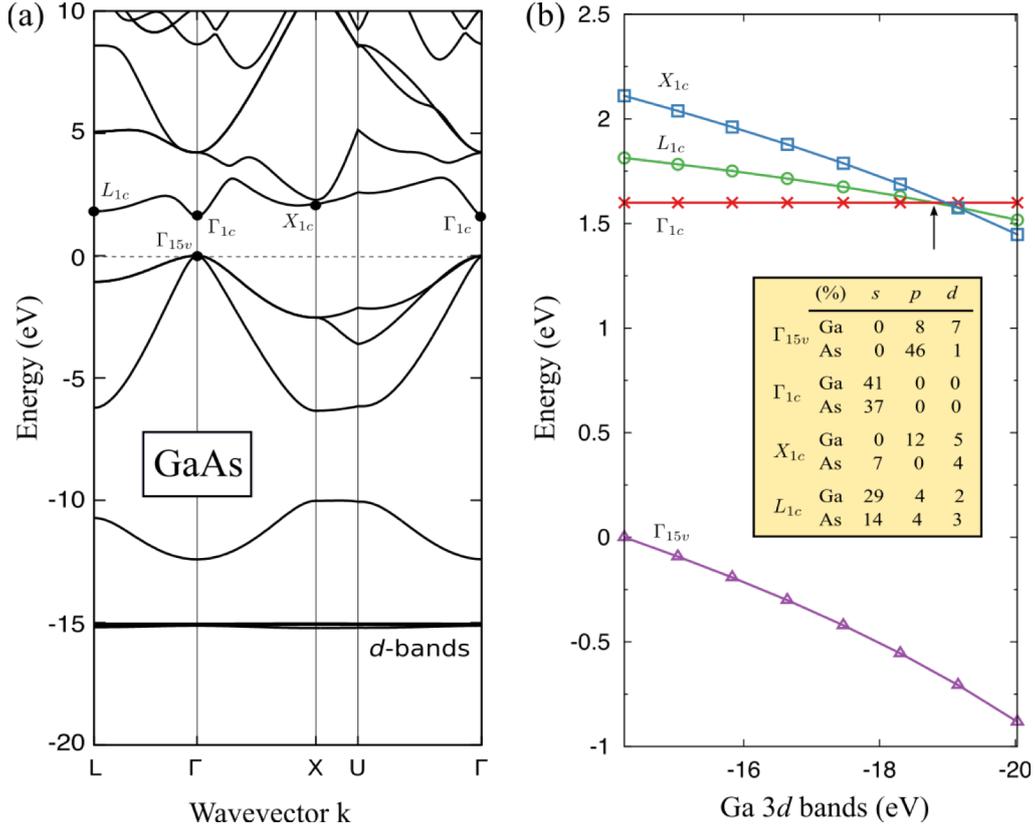

FIG 2. Band structure of GaAs and the varying of its band edge levels as a function of the energy position of the Ga 3$d$ bands. (a) The band structure of GaAs calculated using the mBJ-GGA approach. (b) The varying of the band edge levels of the CB Γ-, X- and L-valley and VBM as pulling the Ga 3$d$ bands down through increasing the applied on-site Coulomb U on the Ga 3$d$ orbitals relying on the mBJ-GGA approach. The vertical arrow indicates the transition between direct and indirect bandgap because of the crossing between Γ-valley and L-valley. Inset to (b) summaries the atomic orbital components of GaAs band edges at U=0.

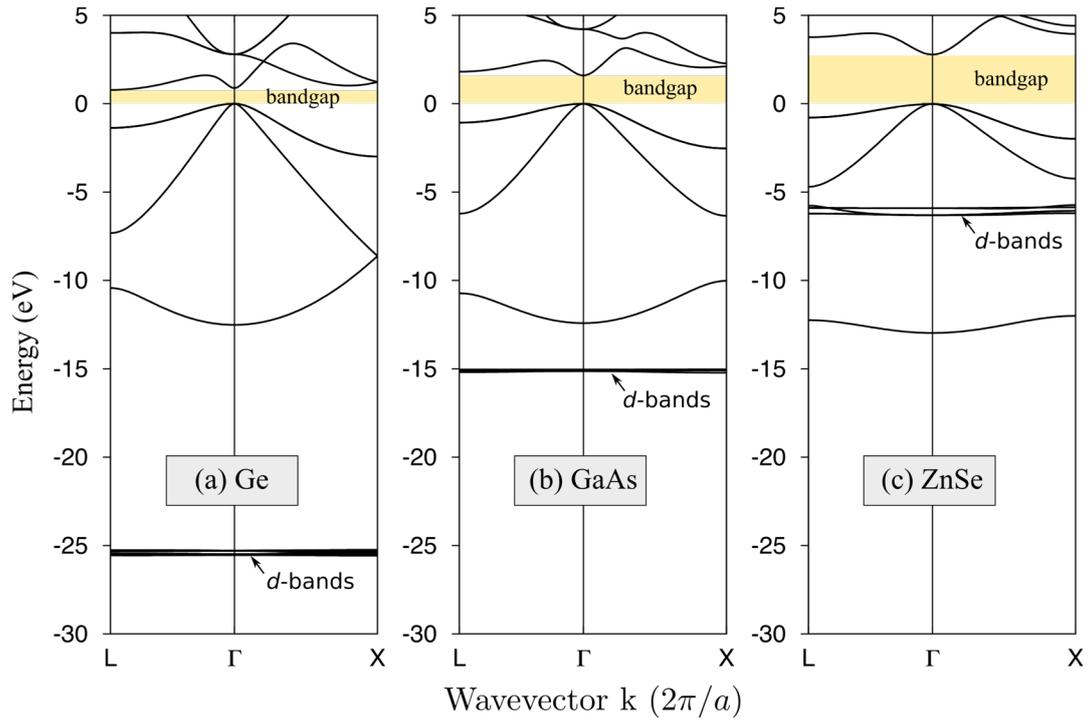

FIG 3. Band structure of (a) Ge, (b) GaAs, and (c) ZnSe predicted by the first-principles mBJ-GGA approach without considering the spin-orbit interaction. Yellow area indicates the bandgap. All energies are relative to the valence band maximum (VBM), which are set to zero. The lattice constants of Ge, GaAs, and ZnSe are very similar.

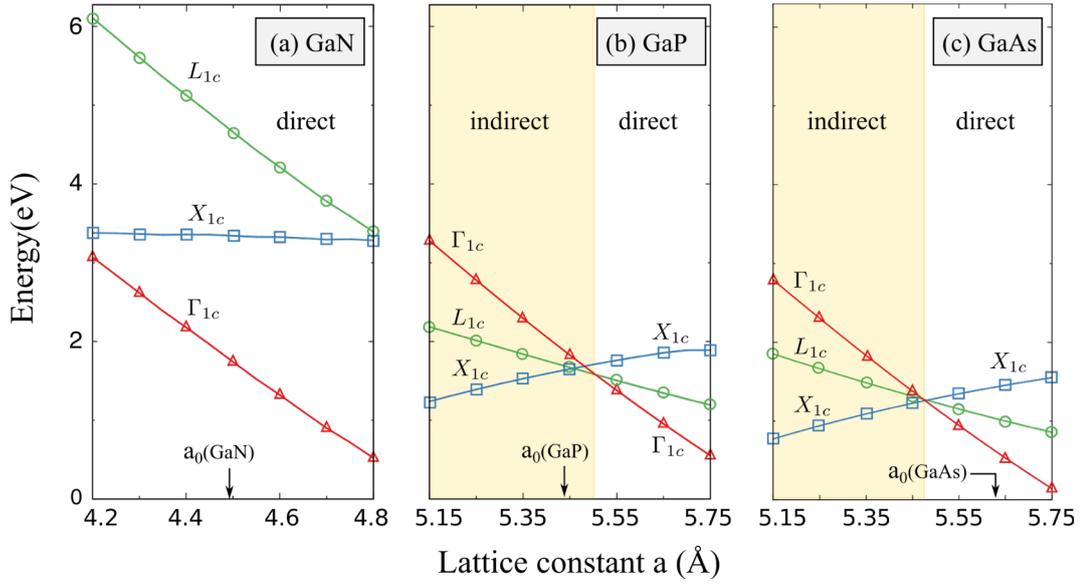

FIG 4. Energies of the CB Γ-, X-, L-valley relative to the VBM as a function of the change of the lattice constant in (a) GaN, (b) GaP, and (c) GaAs. Arrows indicate the experimentally measured lattice constant $a_0$ for each compound. The yellow areas mark the bandgap become indirect.